# Multi databases in Health Care Networks

Nadir K.Salih  Tianyi Zang  Mingrui Sun
School of Computer Science and Engineering, Harbin Institute of Technology, China

*Abstract*

*E-Health is a relatively recent term for healthcare practice supported by electronic processes and communication, dating back to at least 1999. E-Health is greatly impacting on information distribution and availability within the health services, hospitals and to the public. E-health was introduced as the death of telemedicine, because - in the context of a broad availability of medical information systems that can interconnect and communicate - telemedicine will no longer exist as a specific field. The same could also be said for any other traditional field in medical informatics, including information systems and electronic patient records. E-health presents itself as a common name for all such technological fields. In this paper we focuses in multi database by determined some sites and distributed it in Homogenous way. This will be followed by an illustrative example as related works. Finally, the paper concludes with general remarks and a statement of further work.*

## 1. Introduction

The advent of the internet had a major impact on the healthcare industry in the last four decades. While the sophistication of Public Digital Assistant (PDA), wireless systems and browser based technology is at the forefront of all healthcare entities considering implementation and/or expansion of their technology, there are no limits as to how far these will go. With all major financial decisions comes bench marking for best practices, conflicts and negotiations. In health care networks computers are being used with increasing enthusiasm, although the exploitation of their capabilities still lags behind that of industry in general. The 'information technology revolution' has made a great impact on daily life in modern society, being used in familiar applications from high-street cash dispensers to children's education. Increasing exposure to computing tools and services has meant that much of the mystique surrounding the discipline is disappearing, and the next generation of medical professionals and other workers in the health sector can be expected to have a very positive approach to computing [1]. Most of today's Hospital Information Systems (HIS) are characterized by a large number of heterogeneous system components [2].

A multidatabase system consists of a collection of autonomous component database systems. Distribution of data across multiple sites is a clear trend in many emerging internet applications. One major advantage of data distribution is each site can process it's own data with some degree of autonomy and user's can be provided with a single global view of the data[3]. Through Internet, the e-health care could point out record, measure, monitor, manage, and in the end to deliver patient oriented along with condition-specific care services in real time. Internet-based e-health is capable of operating ubiquitously, at anytime for anyone. E-health has been making health care more effective, allowing patients and professionals to do the previously impossible through the widespread information and communication technologies [4]. The telemedicine system, a currently used information system, enabled to maximize the collection, delivery, and communication of health care information, clinical messages, nursing interaction, and medical records from one location to another in e-health fields [6]. Nadir k.Salih et al. [8] They have recommended an agent-Web service that has the features of both the agent technology as well as the Web services technology and is managed by an autonomic system based on multi-agent support. This can help to develop enterprise IT systems that are optimal, highly available. And building deployable solutions in the number of application domains comprising complex, distributed systems.

The remainder of the paper is structured as follows: Section two presents Objectives of computerized information systems in a health network. Section 3 describes some sites and databases in health care networks. Section 4 demonstrates Homogenous Distributed Database Systems. Section 5 Related works, Section 6. Finally, the paper concludes with general summary

## 2. Objectives

Three factors will greatly influence the further development of information processing in health care with in the near future: the development of the population, medical advances, and advances in informatics. Healthcare in the 21$^{st}$ century requires secure and effective information technology systems to meet two of its most significant challenges: improving the quality of care while also controlling the costs of care. The demands of computerized information systems in a health network with regard to hardware and software are rarely matched by industrial applications. The problems to be solved are therefore correspondingly diverse and require many innovative techniques. The objectives of such computerization are to:

• Reduce the need for, and duration of, treatment of patients by good prevention methods and early diagnoses;

• Increase the effectiveness of treatment to the extent allowed by improved information;

• Relieve professionals and other workers in care units of information processing and documentation burdens, thereby freeing them for more direct care of the patient;

• Enhance the exploitation of resources available for health care by good management and administration;

• Archive clinical data, facilitate the compilation of medical statistics and otherwise support research for the diagnosis and treatment of disease.

## 3. Some sites and databases in health care networks

During the data analysis phase of a study of such a typical network, six important sites or functional area types were identified as providing a realistic and representative scenario for a case study. The sites were

(1) Community care units for schools and the community;

(2) General practitioner or health centre units providing a first contact point for patients with the health care network;

(3) Casualty units within hospitals for treating accidents or other emergencies requiring urgent responses;

(4) Laboratories, usually within hospitals, to provide analyses of samples from various other units; (5) Patient records offices, for the administration of medical records within hospitals;

(6) Wards for the treatment of in-patients within hospitals.

The requirements of each of these sites were distilled from the output of the data analysis phase of the study, supplemented by data descriptions reported by other studies, and expressed as a collection of five descriptions: data objects handled, functions carried out, events triggering each function; constraints on the functions; and the distribution of the functions and the data around the network.

## 4. Homogenous Distributed Database Systems

A homogenous distributed database system is a network of two or more databases that reside on one or more machines [7]. Figure 1 illustrates a distributed system that connects three databases: COMMUNITY CARE DATABASE, HEALTH CENTRE DATABASE and CASUALTY. An application can simultaneously access or modify the data in several databases in a single distributed environment. For example, a single query from a COMMUNITY CARE DATABASE client on local database can retrieve joined data from the PATIENT table on the local database and the DOCTOR table on the remote HEALTH CENTRE database. For a client application, the location and platform of the databases are transparent. You can also create synonyms for remote objects in the distributed system so that users can access them with the same syntax as local objects. For example, if you are connected to database COMMUNITY CARE DATABASE yet want to access data on database HEALTH CENTRE DATABASE, creating a synonym on COMMUNITY CARE DATABASE for the remote PATIENT table allows you to issue this query:

SELECT * FROM PATIENT;

In this way, a distributed system gives the appearance of native data access. Users on COMMUNITY CARE DATABASE do not have to know that the data they access resides on remote databases.

## 5. Related works

Mobile multi-agent information Platform MADIP that is developed on top of JADE and allows MAs to work on behalf of health care professionals, to collect distributed users' vital sign data, and to spontaneously inform abnormal situations to associated health care professionals [4]. National Health Information Network (NHIN) This model shows the feasibility of an architecture wherein the requirements of care providers, investigators, and public health authorities are served by a distributed model that grants autonomy, protects privacy, and promotes Participation [5]. Present the design and architecture of a mobile multi-agent based information platform – MADIP – to support the intensive and distributed nature of wide-area (e.g., national or metropolitan) monitoring environment. To exemplify the proposed methodology, an e-health monitoring environment was built on top of MADIP [6].

## 6. The appropriateness of DDB technology for health applications

There is an interesting hierarchy or network of (distributed) databases within the distributed databases for this application. There are many (distributed) databases in the system which corresponds to individual patients, doctors and many other objects. Some of these may appear as simple entities in other distributed databases. Consider an extreme case of a patient who has some chronic condition such as asthma or hypertension, which may be particularly persistent. He or she could quite conceivably also be found, perhaps several years later, to be suffering from some other chronic condition, for example arthritis. A patient like this could, over a decade say, accumulate a sizeable collection of details resulting from frequent consultations, tests and treatments. Now all patients' records constitute a distributed database in a structural or intensional sense. However the physical size of the record of an individual chronically ill patient could mean that it qualifies as a database in the extensional sense also.

## 7. Conclusions

We have looked at the objectives of computerization of health care systems, and we have concentrated up on some sites and databases in health care networks. We also provide an intuitively acceptable set of criteria to help determine if the DDB approach is appropriate for a particular application environment.

We are also working with healthcare professionals to ensure that our e-healthcare system meets their needs, and we are improving our systems based on their feedback.

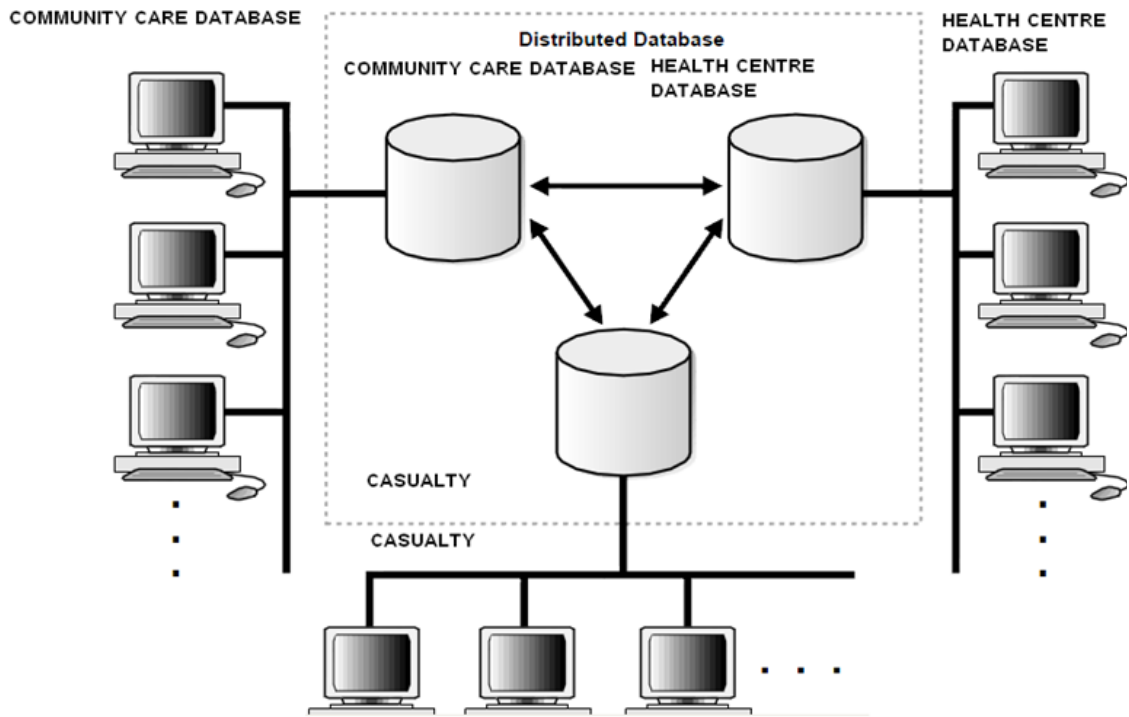

Figure 1 Homogeneous Distributed Database
(Self Creation)